\documentstyle[epsfig]{mn}

\begin{document}

\title[New Be/X-ray pulsars]
    {Discovery of two new persistent Be/X-ray pulsar systems}

\author[Reig \& Roche]
{Pablo Reig$^{1,2}$ and Paul Roche$^3$\\
$^{1}$Foundation for Research and Technology-Hellas. 711 10 Heraklion. Crete. 
Greece. \\
$^{2}$Physics Department. University of Crete. 710 03 Heraklion. Crete. Greece\\
$^{3}$Astronomy Centre. CPES. University of Sussex. BN1 9QJ. UK
}

\date{Accepted \\
Received : Version Date of current version \\
In original form ..}

\maketitle

\begin{abstract} 

We present RXTE observations of two recently identified massive X-ray
binaries. RX J0440.9+4431/BSD 24--491 and RX J1037.5--564/LS 1698 are
confirmed as accreting Be/X-ray systems following the discovery of
X-ray pulsations, with barycentric pulse periods of 202.5$\pm$0.5 s
and 860$\pm$2 s respectively. The X-ray spectral analysis shows that
the energy spectra of the pulsars can be represented by a power-law,
modified at low energy by an absorption component and at high energy
by a cut-off. Very weak Fe lines may be present. Both sources appear to
display a low cut-off energy when compared to typical X-ray pulsars,
low X-ray variability (factor of $\leq$10), and no dependence of the
X-ray spectrum with energy. Given the similarity of these X-ray
properties with those of the other persistent BeXRB pulsars,
4U0352+309/X Per and RX J0146.9+6121/LS I +61 235, we suggest that RX
J0440.9+4431/BSD 24--491 and RX J1037.5--564/LS 1698 are also members
of this subclass.

\end{abstract}


\section{Introduction}

Motch et al.  (1997)  (hereafter M97) reported the discovery of several new
high mass  X-ray  binaries.  These  systems  were  found  during  the ROSAT
galactic  plane survey by  cross-correlating  the position of  low-latitude
X-ray sources ($|b|$ $<$ 20$^{\circ}$) with SIMBAD OB star catalogues.  The
search  was  restricted  to stars  earlier  than B6 and X-ray  luminosities
$\ge$~10$^{31}$  erg  s$^{-1}$.  When the X-ray  data  were  combined  with
optical  observations,  five sources remained  candidates for new accreting
binary  systems.  These sources are BSD 24--491 (RX  J0440.9+4431),  LS 992
(RX  J0812.4--3114), LS 1698 (RX  J1037.5--5647,  also probably  4U1036--56
according  to  M97),  LS  5039  (RX  J1826.2--1450)  and LS I +61  235  (RX
J0146.9+6121).  In this paper we present RXTE  observations  of BSD 24--491
and LS 1698.  Hereafter,  we will  refer to  sources  by the  names  of the
optical companion.

Be/X-ray binaries (BeXRBs) and supergiant systems represent the
general class of High Mass X-ray Binaries.  These are systems
consisting of an early type star, the exact size forming part of the
classification of the system.  In addition to the normal stellar
component, a compact object is also present in the system, and
accretion onto it is the principal source of X-ray emission in the
system.  Follow-up optical spectroscopic observations carried out by
M97 showed H$\alpha$ in emission in both, BSD 24--491 and LS 1698,
hence classing them as Be stars.  The spectral type for both sources
is B0V-IIIe (M97).

A Be star is an early type luminosity class III-V star, which at some
time has shown emission in the Balmer series lines. This emission, as
well as the characteristic infrared excess, is attributed to the
presence of circumstellar material, most likely forming a disc around
the equator of the Be star.

BeXRBs were first identified by the presence of X-ray flares, and so
were named X-ray transients. Type I flares were observed to occur
regularly, with a fixed period, and it was hypothesised that the
systems consist of a neutron star in an eccentric orbit around the Be
star.  This model predicts that an X-ray flare will be observed during the
time of the neutron star's periastron passage, and this explains the
periodicity observed in the timing of the flares. The size of the
flare will be dictated by the amount of material available for
accretion and by the magnetic field of the neutron star. The flare
occurs at periastron when the neutron star impinges on the region of
denser circumstellar material surrounding the Be star. BeXRBs also
show giant X-ray outbursts (Type II events) in which the X-ray
luminosity increases by a factor 100-1000 above the quiescent
level. The X-ray luminosity during the outburst may reach 10$^{38}$
erg s$^{-1}$, close to the Eddington luminosity. These outbursts do
not correlate with the orbital phase, but occur in an unpredictable
way and are thought to be related to mass ejections from the Be star's
circumstellar envelope.

It has been proposed that a subclass of BeXRBs exists which are
characterised by persistent, weak X-ray emission, and which do not
display the type I or II outburst behaviour. This subclass is
currently represented by X Persei and LS I +61 235 (Haberl, Angelini,
Motch and White, 1998). These systems have relatively low X-ray
luminosities ($\sim$10$^{34}$ erg s$^{-1}$) and long spin periods (837
s X Persei and 1412 s LS I +61 235). To date, these two represent the
only examples of persisent BeXRBs amongst the $\sim$40 known BeXRB
systems. For a recent summary of the observed properties of BeXRBs,
see Negueruela (1998).

\section{Observations}

\subsection{X-ray observations}

The sources were observed with the {\em Proportional Counter
Array} (PCA) onboard the {\em Rossi X-ray Timing Explorer} (RXTE) in
1998 February.  Table 1 shows the journal of the X-ray
observations.  The total on-source time was 20 ks for each target.
The PCA covers the lower part of the energy range, 2-60 keV, and
consists of five identical coaligned gas-filled proportional units
(PCU), providing a total collecting area of $\sim$ 6500 cm$^2$, an
energy resolution of $<$ 18 \% at 6 keV and a maximum time resolution
of 1$\mu$s.  For a more comprehensive description of the RXTE PCA see
Jahoda et al. (1996).

Good time intervals were defined by removing data taken at low Earth
elevation angle ($<$ 8$^{\circ}$) and during times of high particle
background. An offset of only 0.02$^{\circ}$ between the source
position and the pointing of the satellite was allowed, to ensure that
any possible short stretch of slew data at the beginning and/or end of
the observation was removed. This screening criteria allowed us to
divide the observations up into continuous sections of clean data, on
which the X-ray analyses were carried out. The main objective of these 
observations was to search for pulsations in the 
X-ray flux in an attempt to confirm the accreting nature of the sources.


\begin{table*}
\begin{center}
\label{tabx}
\caption{Journal of the X-ray observations}
\begin{tabular}{lcccccc}
\hline
Name of   & Date & Start time & Stop time  & Luminosity$^{a,b}$ &Spin Period & 
HR$^c$\\
Source    &            & 	&     &(erg s$^{-1}$)	& (s)  &\\
\hline 
BSD 24--491 &30/01/98 &06:24:25 &11:19:14  &3.0$\times$10$^{34}$& 202.5$\pm$0.5 
& 
--0.56\\
	   &01/02/98 &05:26:38 &11:07:14  &			&&\\
LS 1698	   &04/02/98 &16:25:14 &19:49:14  &4.5$\times$10$^{35}$ 
&860$\pm$2&--0.40 
\\
	   &05/02/98 &08:40:50 &09:20:14  &			&&\\
	   &05/02/98 &10:03:38 &16:35:14  &			&&\\
\hline
\end{tabular}
\end{center}
{\small a: in the energy range 3-30 keV} \\
{\small b: The assumed distances are (from M97) BSD 24--491: 3.2 kpc, LS 1698: 
5.0 kpc} \\
{\small c: HR=(10-30)--(3-10)/(3-30)}
\end{table*}

\section{Timing analysis}

In this section we present the results of the X-ray  timing  analysis.  For
each source  lightcurves,  pulse profiles and hardness ratios are given and
the best value of the spin period determined.

\subsection{BSD24-491}

\begin{figure}
    \begin{center}
    \leavevmode
\epsfig{file=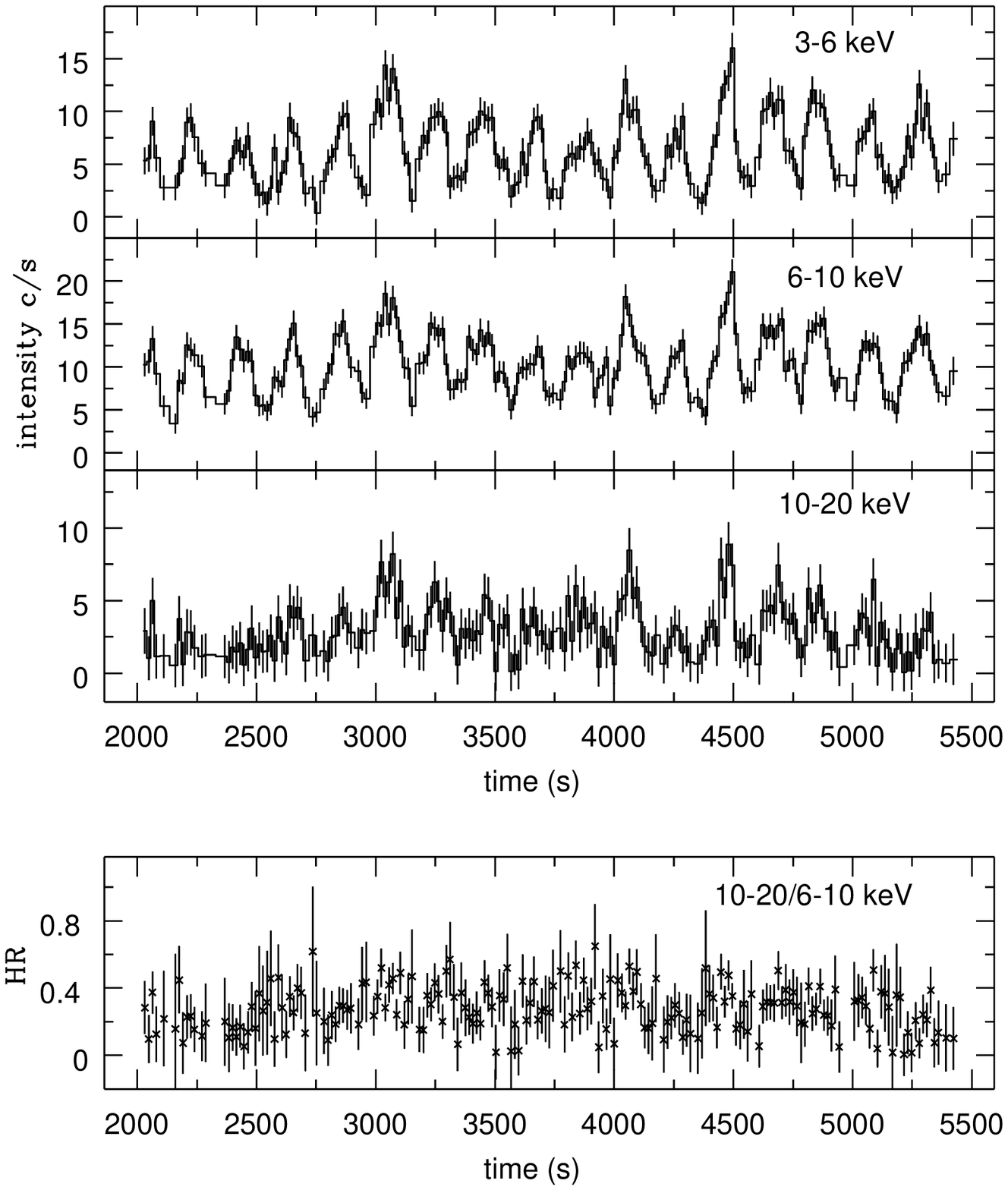, width=8cm, bbllx=50pt, bblly=160pt,
  bburx=525pt, bbury=715pt, clip=}
 \end{center}
        \caption{X-ray pulsations in BSD24--491 at three different energies, 
with each bin representing 16s. The starting time is JD 2,450,843.398. Also 
shown is the variation of the HR 10-20/6-10 keV with time}
        \label{bsd_lc}
        \end{figure}

Fig  \ref{bsd_lc}  shows an  enlarged  section of the  lightcurve  at three
different  energy  ranges:  3-6, 6-10, 10-20 keV.  Pulsations  are  clearly
seen at the three energy  ranges  considered  and they show up in the Power
Density  Spectrum (PDS) as a peak at $\approx$ 0.005 Hz, which  corresponds
to a spin period of $\sim$ 200 s.  The X-ray intensity of the source, after
background subtraction, was 10.7$\pm$0.1 PCA c s$^{-1}$ in the energy range
3-30 keV.  Other than  pulsations,  no other type of  variability  was seen
during the $\sim$ 6 hour  interval  of our  observation.  The  extrapolated
luminosity in the energy range  0.1-2.4 keV is 1.4 $\times$  10$^{33}$  erg
s$^{-1}$,  compared to the  maximum  value of 6.0  $\times$  10$^{33}$  erg
s$^{-1}$ given by M97 during a ROSAT pointed  observation,  indicating some
long term  variability.  In fact M97 reported a variation in the count rate
of a factor of 2.5 over their observations.

In order to  determine  the  pulsation  period,  a period  search  by epoch
folding  was  performed  near to the  period  expected  from the FFT  power
spectrum.  A solar barycentric  pulse period of 202 s was obtained from the
resulting pulse-folding  periodogram.  The pulse profile obtained from this
period  was  used as a  template.  Then,  both  the  observations  and  the
template were divided into 4 roughly equal time  intervals.  The difference
between the actual pulse period at the epoch of observation  and the period
used to fold the data was  determined  by  cross-correlating  the  original
light curve and the  template.  The  $P_{actual}-P_{template}$  shifts were
fitted to a linear function to obtain an improved period.  A new template
was derived from this period and the process  repeated.  The  derived  pulse
period was 202.5$\pm$0.5 s.  The error represents the scatter of the points
about the best-fit straight line.  The accuracy in the determination of the
spin period is limited by the length of the time series with respect to the
length of the period.

\begin{figure}
    \begin{center}
    \leavevmode
\epsfig{file=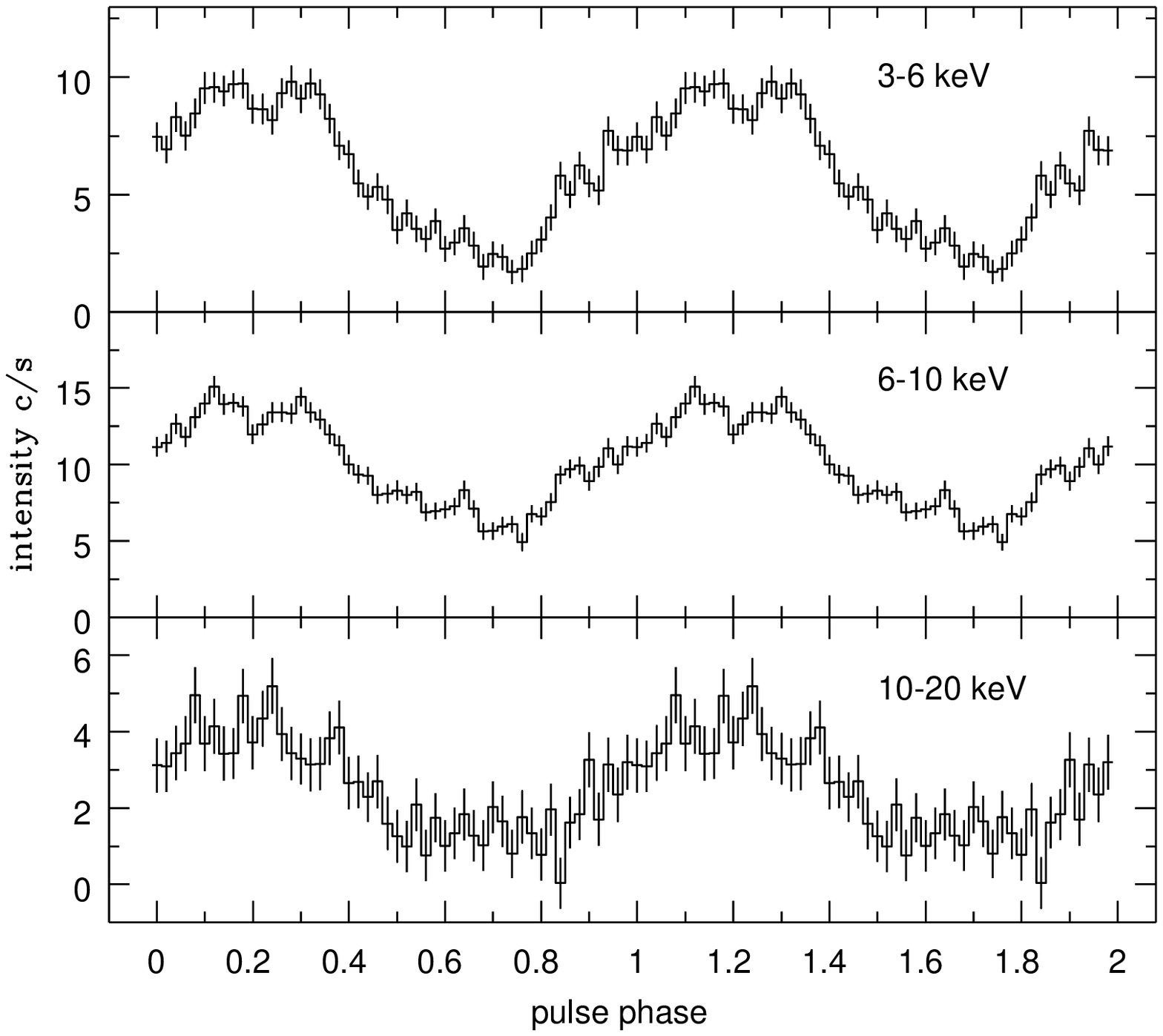, width=8cm, bbllx=60pt, bblly=285pt,
  bburx=540pt, bbury=710pt, clip=}
 \end{center}
        \caption{Pulse profiles of BSD 24--491 at three different energy bands}
        \label{bsd_pp}
        \end{figure}

Once the pulse period is known, the pulse profile can be derived by
folding the lightcurve modulo the best fit value of the pulse period.
Such a pulse profile is shown in Fig \ref{bsd_pp} at the same three
energy ranges in which the lightcurve was separated.  No depence on
energy is seen.  The pulse shape is nearly sinusoidal, as might be
expected from the absence of second or higher harmonics in the
PDS. The amplitude of the modulation is $\sim$ 75\% in the 3-30 keV
energy range.


\subsection{LS 1698}

M97 reported a maximum X-ray luminosity  L$_x$(0.1--2.4  keV) $\approx$ 1.1
$\times$ 10$^{34}$ erg s$^{-1}$ for LS 1698.  The average 3-30 keV RXTE PCA
count rate, after background  subtraction, was 45.8$\pm$0.1 count s$^{-1}$,
without  significant  changes  in  intensity.  The PDS  averaged  over  two
intervals of 3600 s revealed the  presence of one peak at $\nu  \sim$0.0011
Hz, corresponding to a coherent modulation with a period of $\sim$900 s.

\begin{figure}
    \begin{center}
    \leavevmode
\epsfig{file=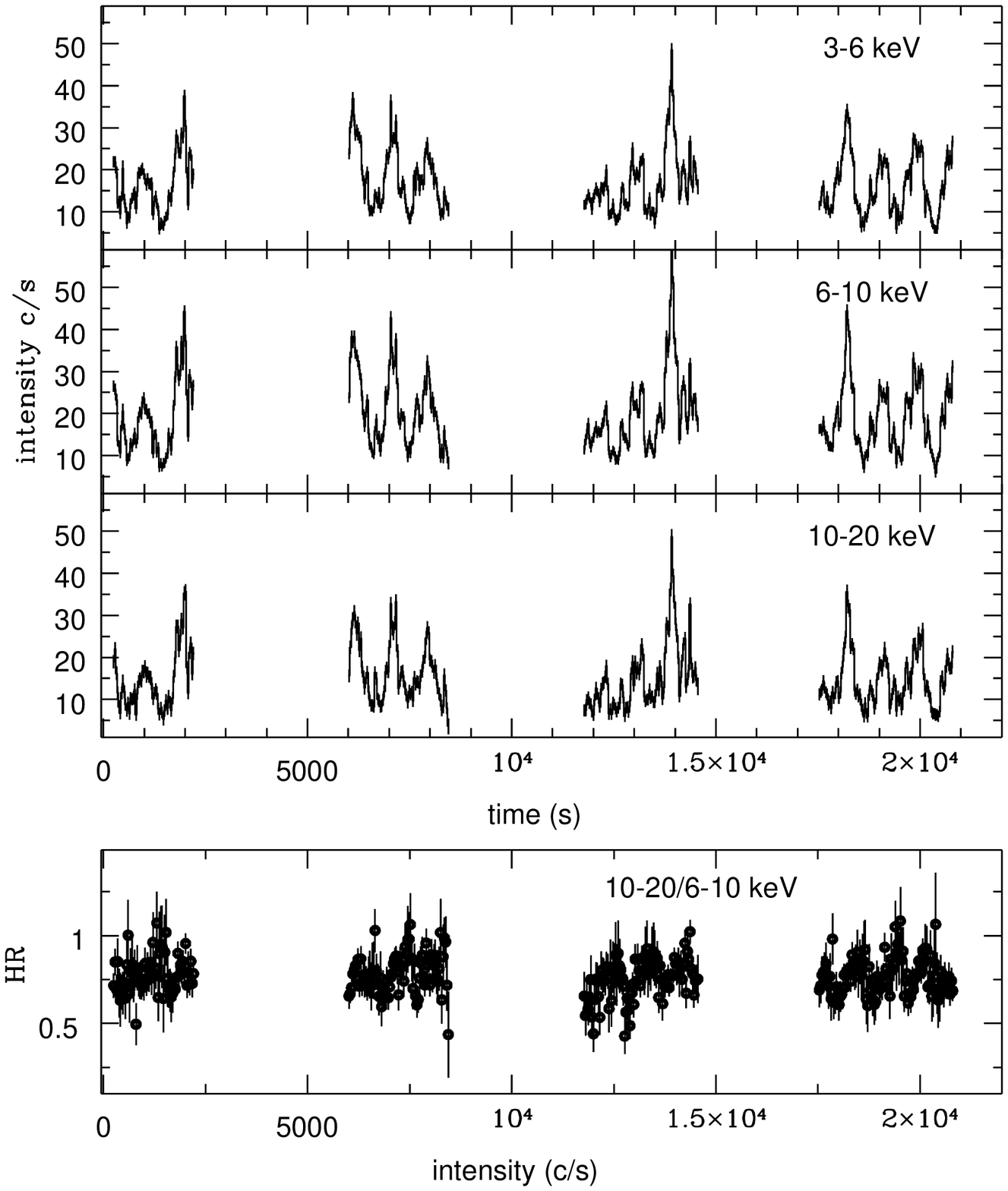, width=8cm, bbllx=35pt, bblly=80pt,
  bburx=540pt, bbury=715pt, clip=}
 \end{center}
        \caption{X-ray lightcurve of LS 1698 showing pulsations at different 
energy bands. Each bin represents 32 s. Time 0 is JD 2,450,849.420. Also shown 
is the hardness ratio 10-20/6-10 keV as a function of time. Note that the HR 
follows the pulses.}
        \label{lc1698}
        \end{figure}

\begin{figure}
    \begin{center}
    \leavevmode
\epsfig{file=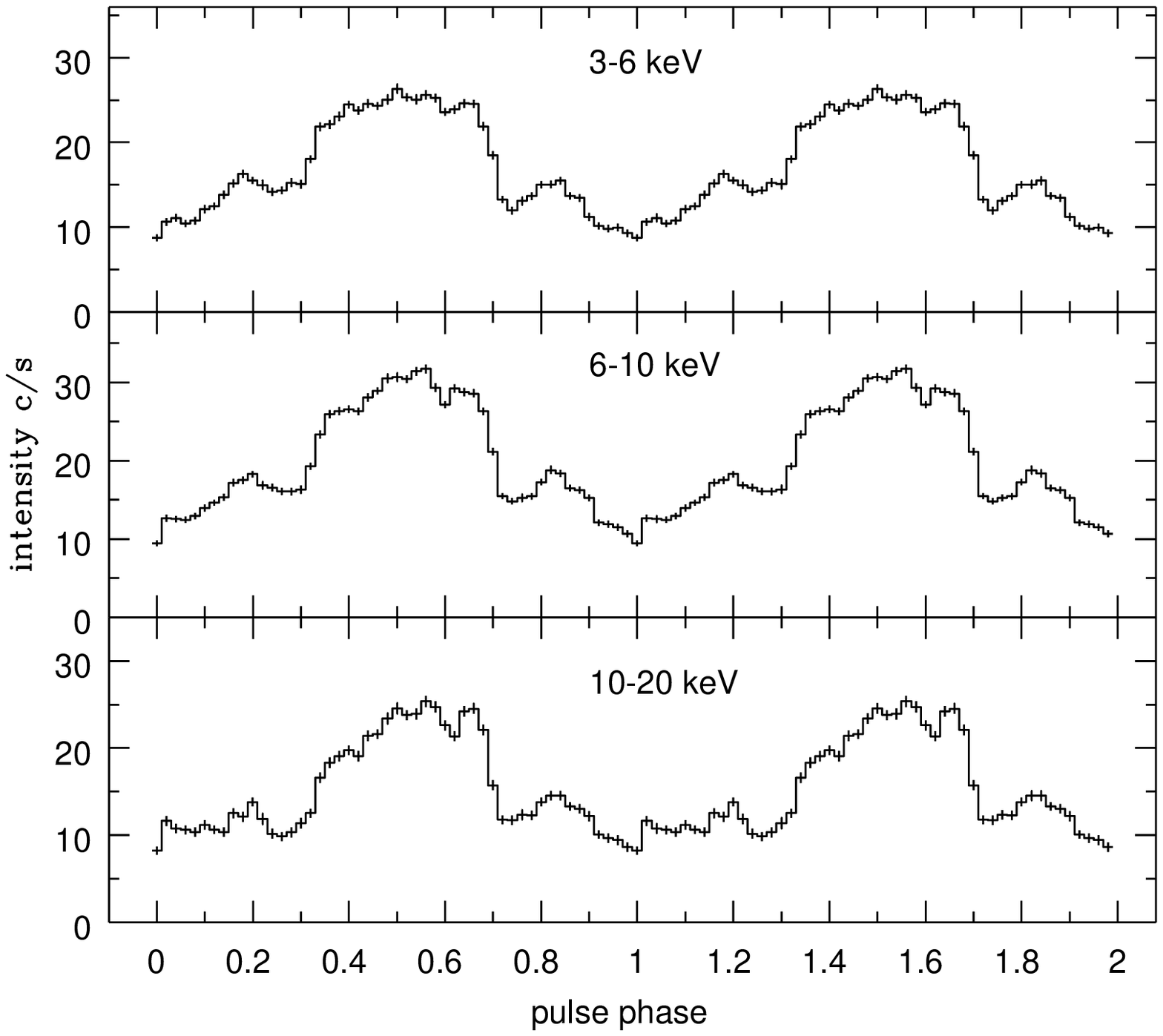, width=8cm, bbllx=60pt, bblly=295pt,
  bburx=540pt, bbury=711pt, clip=}
 \end{center}
        \caption{Pulse profiles of LS 1698 at 3-6, 6-10 and 10-20 keV.}
        \label{pp1698}
        \end{figure}

In order to determine a more accurate value of the pulse period we rebinned
the  lightcurve  into  32  s  bins  spanning  a  total  on-source  time  of
$\sim23500$s  (including gaps due to Earth  occultations or passage through
the South  Atlantic  Anomaly).  Times were  corrected  to the Solar  System
barycentre, and periodicities  searched for in these lightcurves  using the
epoch  folding  technique.  Data were folded over a range of trial  periods
and the $\chi^2$ of the folded  lightcurve  calculated.  The period adopted
is the one which shows a maximum in the  $\chi^2$  versus  period  diagram.
The period  providing  the highest  $\chi^2$ was found to be  860$\pm$2  s.
Although the  observations  do not cover a long enough base line to further
constrain  this  period,  e.g.  by applying  the  cross-correlation  method
explained  above,  pulsations are distinctly  seen in the X-ray  lightcurve
(Fig \ref{lc1698}).  Fig \ref{pp1698}  displays the 3-6, 6-10 and 10-20 keV
folded  lightcurves of LS 1698.  These pulse profiles are highly structured
with  a  broad  peak   covering   phases   0.1-0.7.  The  pulse   fraction,
PF=(I$_{max}$--I$_{min}$)/(I$_{max}$+I$_{min}$)   is  $\sim$53$\pm$2\%   and
remains the same for the three  energy  ranges  considered.  I$_{max}$  and
I$_{min}$ are the intensity at pulse peak and pulse dip, respectively.

M97 noted that the source 4U1036--56/3A1036--565 may be the same as RX
J1037.5--564/LS 1698, based on the positional coincedence of the ROSAT error
circle with those of Uhuru and Ariel V.  However, despite observations with
several satellites (Uhuru, Ariel V, OSO 7, EXOSAT, ROSAT), no pulsations have
previously been detected from 4U1036--56 (although this may simply be
attributable to the lower sensitivity of previous observations rather than a
real indication that no pulsations are present).  The reported detection of a
flare in the Ariel V observations (Markert et al.  1979) only shows a slight
increase ($\times$2.4) over the Uhuru flux in the same energy region (2-10 keV),
and the ROSAT observations were about an order of magnitude lower than this.
The Uhuru and Ariel luminosities are consistent with those we obtain from our
RXTE observations, $\sim$3 $\times$ 10$^{35}$ erg s$^{-1}$ compared to our value
of $\sim$2 $\times$ 10$^{35}$ erg s$^{-1}$ (2--10 keV and assuming a distance of
$\sim$5 kpc).  It would thus appear that 4U1036--56 displays a low level of
X-ray variability, which might support the identification with LS 1698, but we
cannot confirm this.


\subsection{Hardness ratios}
\label{HR}

\begin{figure}
    \begin{center}
    \leavevmode
\epsfig{file=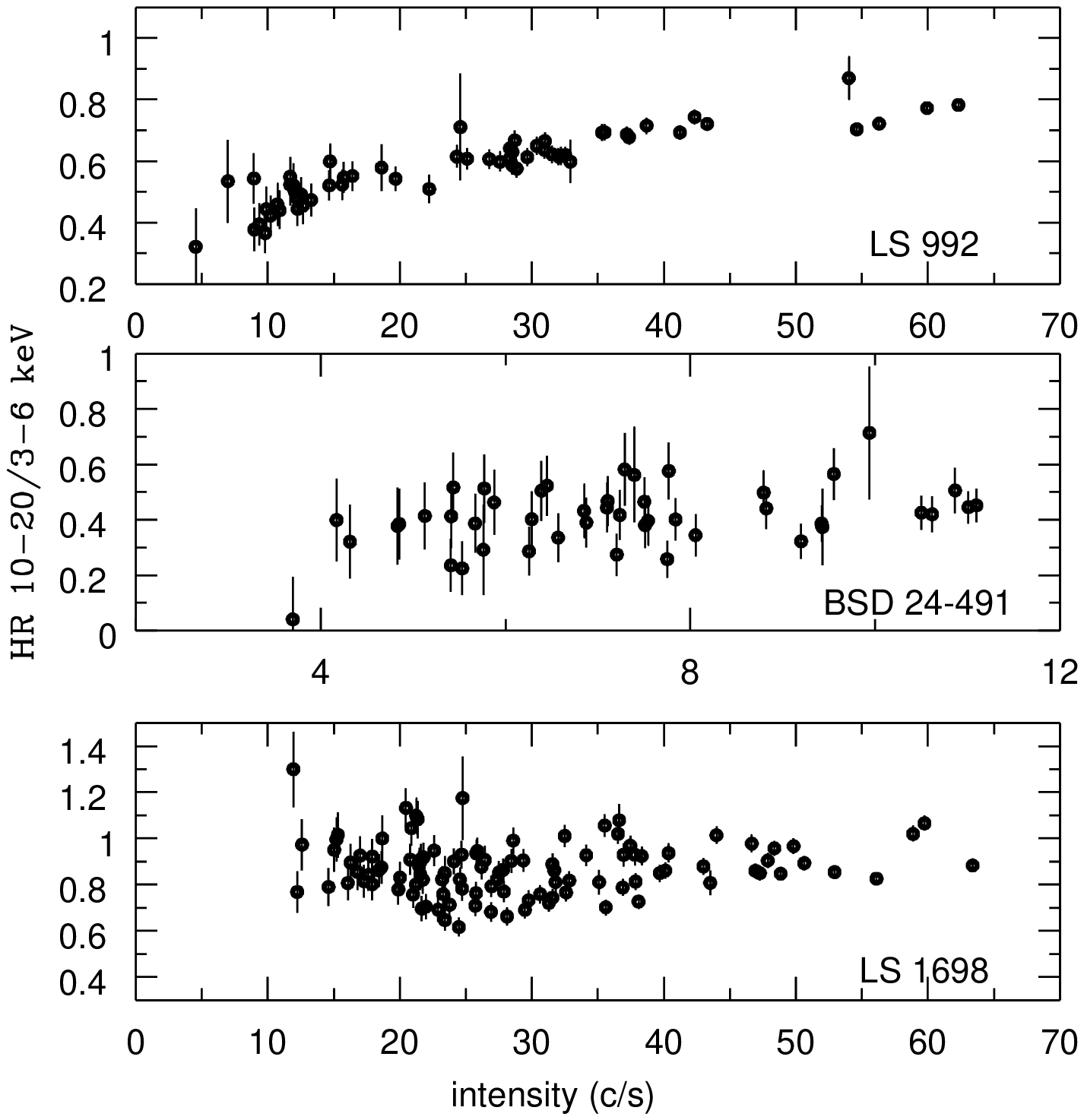, width=8cm, bbllx=80pt, bblly=270pt,
  bburx=510pt, bbury=713pt, clip=}
 \end{center}
        \caption{Hardness ratio 10-20/3-6 keV as a function of intensity of the 
persistent BeXRBs BSD 24--491 and LS 1698. Shown also for comparison is the HR 
variation of a transient BeXRB (from Reig \& Roche, 1999)}
        \label{hr_i}
        \end{figure}

LS 1698 is the hardest of the two sources, as indicated by a hardness ratio
(HR) of --0.40,  defined  as  $(10-30)-(3-10)/(3-30)$,  compared  to --0.56
displayed by BSD 24--491 (see Table 1).  This fact also becomes apparent in
Fig \ref{hr_i}, which shows tha hardness ratios in the photon-energy  bands
10-20 and 3-6 keV as a function of the summed  count rate of the two bands.
It is worth pointing out a few interesting results.  Firstly, the HR of BSD
24--491 and LS 1698 remains  basically  constant at $\sim$0.4 and $\sim$0.9
respectively.  This  contrasts  with the behaviour of the  `typical'  BeXRB
pulsar transient system RX J0812--3114/LS992  shown for comparison, where a
clear  dependence  on intensity is seen (Reig \& Roche 1999).  Secondly, as
stated above, the X-ray  emission  from LS 1698 is much harder than that of
BSD 24--491.

\section{Spectral Analysis}

Although each source adds its peculiarities, the X-ray photon energy
spectra of accreting X-ray pulsars share the following general
characteristics:

\begin{itemize}
	
\item [--] broad-band emission, not dominated by sharp spectral
features 

\item [--] a continuum described in terms of a power-law with low
energy absorption and a cut-off above 10-20 keV 

\item [--] the X-ray flux is mostly emitted in the energy range 2-20 keV 
with a rapid falloff above 20 keV

\item [--] an iron line at $\sim$6.4 keV due to fluorescent
reprocessing by cold circumstellar material is present in a number of
systems

\end{itemize}

In order to investigate the X-ray energy emission of our sources,
fits using a variety of models were performed. Table 2
gives the results of the spectral analysis. We have included only the
values of the models which gave acceptable fits, i.e. reduced $\chi^2$
$\leq$ 2.

\subsection{BSD 24--491}

\begin{figure}
    \begin{center}
    \leavevmode
    \epsfig{file=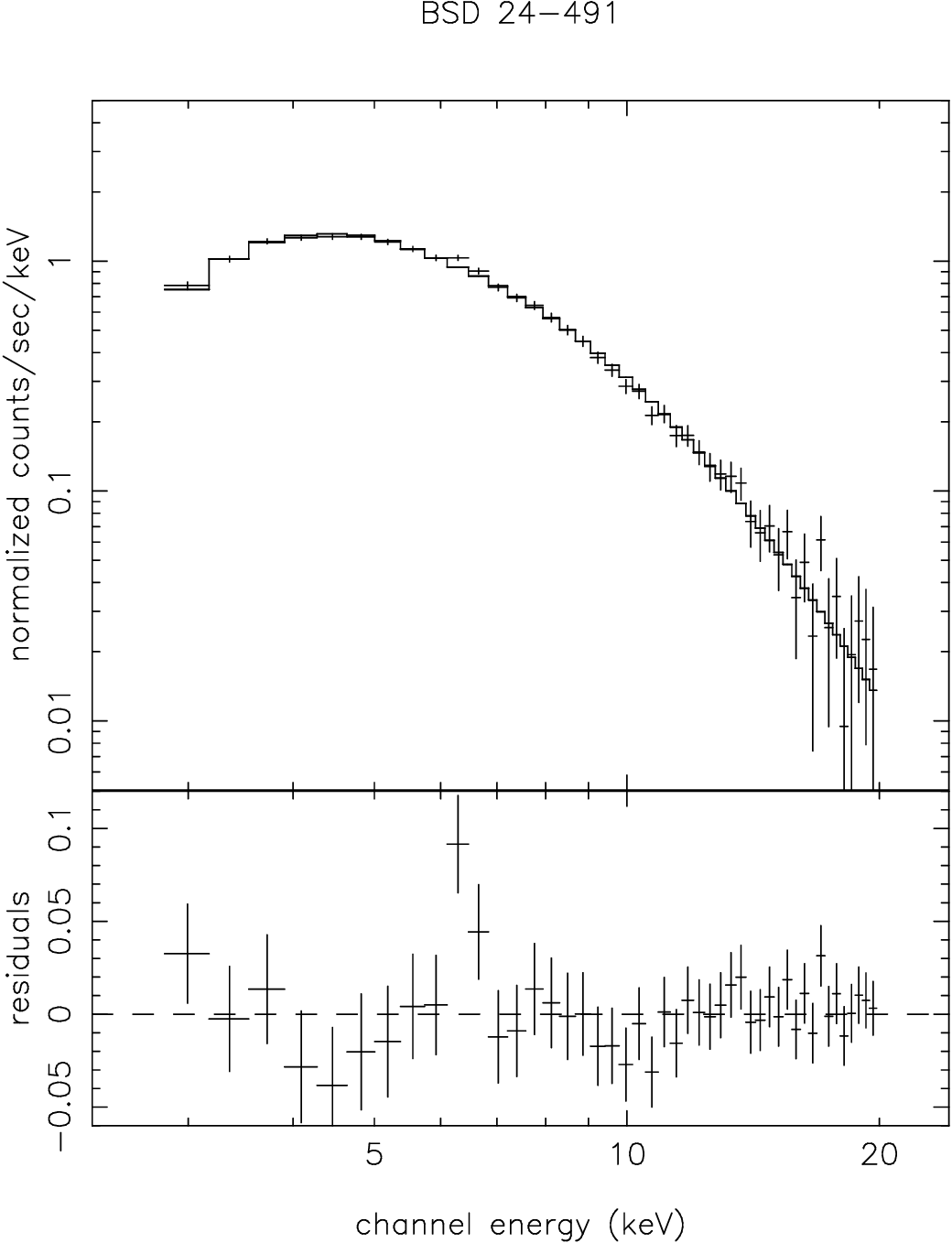, width=7cm, bbllx=34pt, bblly=55pt,
  bburx=540pt, bbury=665pt, clip=}
 \end{center}
        \caption{PCA spectrum of BSD 24--491. The continuum was fit to a 
power-law plus a high energy cut-off. If no Gaussian component is added to the 
fit, an emision feature at $\approx$ 6.2 keV shows up in the residuals}
        \label{sp24}
        \end{figure}

The analysis of the X-ray spectrum of BSD 24--491 is complicated by the low
count  rate.  To improve  the signal to noise  ratio we used only data from
the top layer  anodes of the four  PCUs  that  were on  during  the  entire
observation.  Also, models were fitted to the energy range 3-20 keV instead
of the 3-30 keV used for LS 1698.  In spite of this,  the
errors are too large to  meaningfully  distinguish  between  the  different
spectral  models  listed in Table~2.  An F-test  shows that the
inclusion of a blackbody  component to the power-law model is not justified
by the  reduction  of the  $\chi^2$.  For  the  purpose  of the  subsequent
discussion,  and in order to  compare  spectral  parameters  with the other
sources, we will consider only the results from the power-law  plus cut-off
model.  An  emission  feature at around  6.2 keV shows up in the  residuals
(Fig  \ref{sp24}) but the fit does not allow us to constrain its parameters
precisely.  We set an upper  limit  for the  equivalent  width of the iron
line of  $\sim$  100 eV.  The line  centre  remained  at the same  value of
6.2$\pm$0.2 keV despite fixing the line width to $\sigma \sim$ 0.1, 0.2 and
0.3 keV,  indicating  that the line  width is  narrower  than the  spectral
resolution  of the PCA (18\% at 6 keV).  BSD 24--491  also shows a very low
cut-off  energy of  $\sim$1.9  keV.  This  value is  outside  the  detector
response range and is thus not well constrained.  However, if the iron line
is included in the fit the cut-off energy  increases to $\sim$4.5  keV.  In
the energy range 3-30 keV, L$_x$ $\sim$3.0 $\times$ 10$^{34}$ erg s$^{-1}$,
assuming a distance of 3.2 kpc.

\subsection{LS 1698}

The  best  fit to the  X-ray  spectrum  of LS 1698 is  obtained  using  the
power-law plus high energy cut-off model ($\chi^2_r$=1.23 for 55 dof).  The
inclusion of a Gaussian component,  representing an iron line, improved the
fitting  by  reducing  $\chi_r^2$  to 1.00 for 52 dof.  However,  an F-test
shows that the probability of this happening by chance is higher than 10\%.
The best-fit line centre energy and width are 6.5$\pm$0.2 keV and $\sigma$ 
$\sim$0.2 respectively.  If the Gaussian  component  is
taken into  account,  the  cut-off  energy is 4.7 keV, much lower  than the
values of 10-20 keV typically found in X-ray pulsars.  This may represent a
signature of the persistent BeXRB systems (see below).  An upper limit
for the equivalent width of the iron line is estimated to be $\sim$65 eV.
The  photon  index  lies in the  interval  $\alpha$=1.0--1.2  depending  on
whether  the  iron  line  is  included  in the  fit or  not.  LS  1698  has
E(B-V)$\approx$0.75  (M97),  implying N$_H$  $\sim$0.5  $\times$  10$^{22}$
cm$^{-2}$  (Ryter,  Ceasarsky \& Audouze,  1975).  However, the  absorption
implied  from  the  spectral  fitting  is  higher,  $\sim$4.5  $\times$
10$^{22}$  cm$^{-2}$.  Thus, some amount of matter must be located close to
the system,  which would  justify the presence of the iron line.  The X-ray
luminosity is $\sim$4.5 $\times$ 10$^{35}$ erg s$^{-1}$ in the energy range
3-30 keV and assuming a distance of 5 kpc.

\begin{table*}
\begin{center}
\label{models}
\caption{Spectral fits results. Uncertainties are given 90\% confidence for 
one parameter of interest. Spectra were fitted in the energy range
3-30 keV for LS 1698 and 3-20 keV for BSD 24--491.}
\begin{tabular}{lcc}
\hline
		&  BSD 24--491	& LS 1698\\
\hline
\multicolumn{3}{l}{{\bf Power-law}}\\
$\alpha$	& 2.33$\pm$0.07 & \\
N$_H$ (10$^{22}$ atoms cm$^{-2}$) & 6.2$\pm$0.7\\
$\chi^2_r$(dof) & 1.14(43) & \\

\hline
\multicolumn{3}{l}{{\bf Power-law \& blackbody}}\\ 
$\alpha$	&& 1.84$\pm$0.06\\
kT (keV)	&& 2.9$\pm$0.2\\
R (km)		&& 0.10$\pm$0.01\\
N$_H$ (10$^{22}$ atoms cm$^{-2}$) && 8.2$\pm$0.5\\
$\chi^2_r$(dof) && 1.66(54) \\

\hline
\multicolumn{3}{l}{{\bf Two blackbody}}\\
kT$_1$ (keV)	& 1.30$\pm$0.06 & 1.55$\pm$0.04\\
R$_1$ (km)	& 0.22$\pm$0.01& 0.49$\pm$0.02\\
kT$_2$ (keV)	& 3.8$\pm$0.6 & 4.48$\pm$0.12\\
R$_2$ (km)	& 0.03$\pm$0.01 & 0.085$\pm$0.004\\
N$_H$ (10$^{22}$ atoms cm$^{-2}$) & -            & 2.1$\pm$0.3\\
$\chi^2_r$(dof) & 0.78(43)& 1.34(55)\\

\hline
\multicolumn{3}{l}{{\bf Cut-off power-law *}}\\
$\alpha$	& 1.5$\pm$0.4 & 1.20$\pm$0.08[1.02$\pm$0.07] \\
E$_{cut}$	& $\sim$ 1.9[$\sim$4.5]& 6.2$\pm$0.4[4.7$\pm$0.4]\\
N$_H$ (10$^{22}$ atoms cm$^{-2}$) & 4.1$\pm$1.0[$\sim$1.9] & 
5.5$\pm$0.5[4.5$\pm$0.4] \\
$\chi^2_r$(dof) & 0.94(41)[0.56(40)] &1.23(55)[0.98(53)]\\
\hline
\end{tabular}
\end{center}

{\bf *} The values in [ ] represent the best-fit parameters when an
iron line is included. For BSD 24--491, E$_c$=6.2$\pm$0.2 keV with
fixed $\sigma$=0.1 keV, whereas for LS 1698 E$_c$=6.5$\pm$0.2 keV and
$\sigma \approx$ 0.2 keV
\end{table*}

        \begin{figure}
    \begin{center}
    \leavevmode
    \epsfig{file=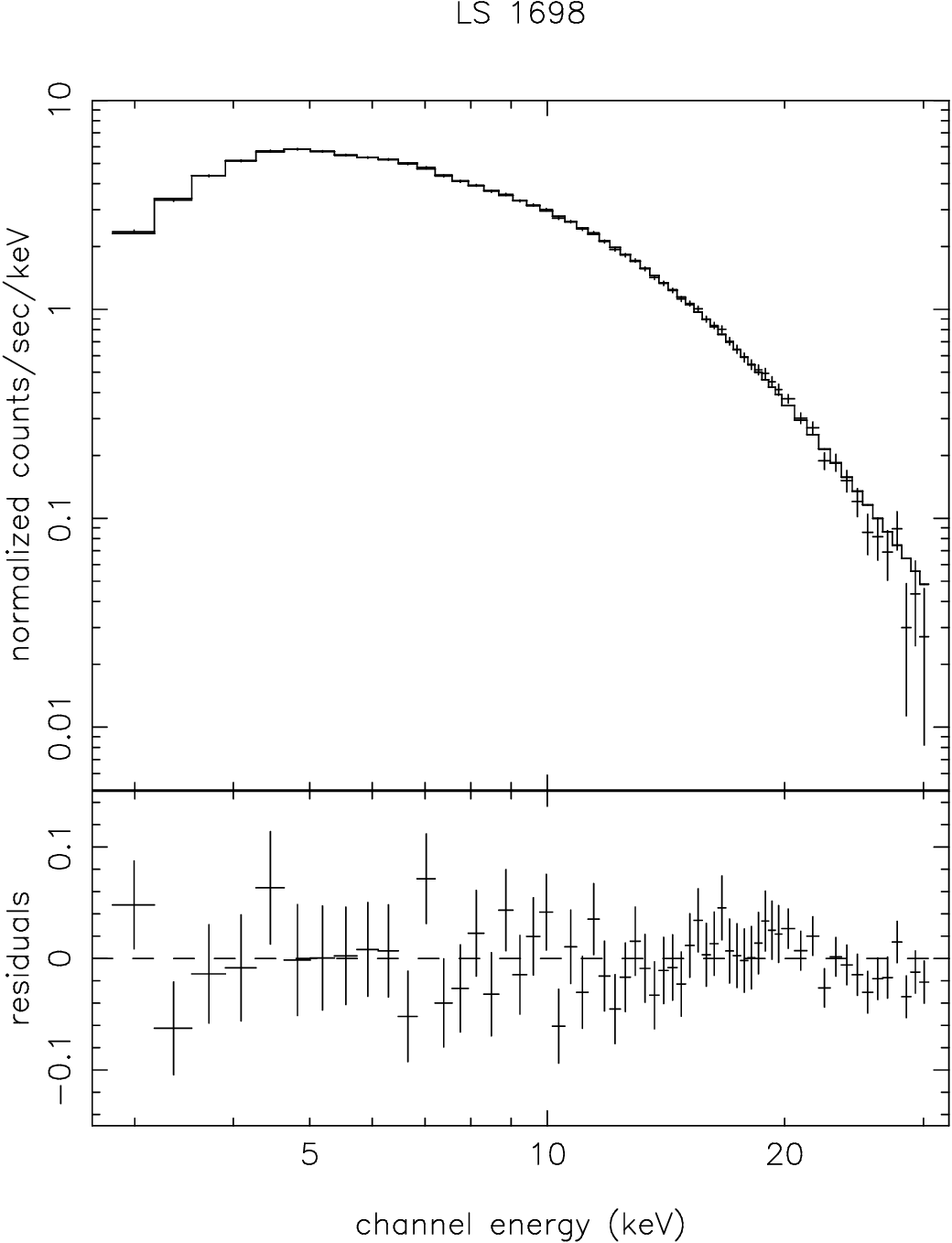, width=7cm, bbllx=34pt, bblly=55pt,
  bburx=540pt, bbury=665pt, clip=}
 \end{center}
        \caption{PCA spectrum of LS 1698 and residuals. The best-fit power-law 
with high energy cut-off is plotted as a straight line}
        \label{sp1698}
        \end{figure}

\section{Discussion}

Pulsations have previously been found in both high and low mass X-ray
binaries.  However, of the 38 X-ray pulsars currently known, only 5
have been identified as LMXRBs, whereas 33 have an OB-type star as the
optical counterpart.  For the confirmed BeXRB systems, around 70$\%$
are found to be pulsators, rising to 75\% if suspected BeXRBs are
included.

The periods of X-ray pulsars are distributed over a factor
$\sim$10$^5$, from 2.5 ms (SAX J1808--369, the bursting millisecond
X-ray pulsar) to 1412 s (LS\,I +61 235, a persistent BeXRB), with no
evidence for clustering at any particular period. The range for BeXRBs
is almost as extensive, the fastest being 69 ms (A 0535-668). 

We have carried out X-ray timing and spectral analyses of two of the
four newly discovered high mass X-ray binaries by M97. We have
detected pulsations in both BSD 24--491 (period 202.5 s) and LS 1698
($\sim$860 s). In addition to slow pulsars, LS 1698 and BSD 24--491
seem to share further similarities in their X-ray variability
characteristics. They have both been detected every time that they
have been observed (during the ROSAT all-sky survey and subsequent
follow-up pointed observations using ROSAT (M97) and here with RXTE)
and they do not appear to show short-term variability (of time scales
of hours or days).

\subsection{Is there a class of persistent Be/X-ray binaries?}

Until recently, X Per/4U0352+309 was unique among massive X-ray binary
systems in that it was the only persistent BeXRB.  The optical
component is a B0Ve star at a distance of $\sim$700 pc (Lyubimkov et
al. 1997).  Persistent high mass X-ray binaries do exist but they
harbour an evolved, i.e.  supergiant, companion.  In these systems the
X-rays are attributed to constant accretion from the strong stellar
wind of the primary.

In the case of X Per, the weak but persistent emission is most likely
a result of the fact that the neutron star is {\em always} a long way
from the Be star and never enters the denser inner regions of the
circumstellar disc. An orbital period of 580 days has been proposed
(Hutchings et al. 1974), but never subsequently confirmed by later
studies. However, there appear to be characteristic timescales of
photometric variability ranging from hundreds to thousands of days
(Roche et al. 1997), but nothing that appears to be orbital in
nature. In most respects, X Per behaves like a classical Be star,
presumably because the separation of the components means that the B0V
star and it's disc are unaffected by the distant neutron star. The
neutron star can be affected by episodes of mass ejection from the
circumstellar disc (e.g. Roche et al. 1993, 1997), but is usually only
accreting from the Be star wind.

\begin{table*}
\begin{center}
\label{persis}
\caption{Comparison of the X-ray properties of persistent and transient BeXRBs}
\begin{tabular}{lllcccccl}
\hline
X-ray & Optical	& Spectral &Spin &Spectral &E$_{cut}$  &L$^b_{\rm x,max}$ &X-ray 
&EW(Fe)\\
name & name &type &period (s) &index &(keV) &$\times$ 10$^{36}$&variability$^c$ 
& 
(eV)\\
\hline
4U0352+309 & X Per	    &B0Ve   & 837   &0.8$\pm$0.2	&2.2$\pm$0.4	
& 0.03 & 5      
& $\leq$ 100 \\
RX J0146.9+6121 & LS I +61 235&B1Ve   &1404   &1.2$\pm$0.2	&4.0$\pm$0.7	
& 0.5  & 20     
& $\leq$ 90 \\
RX J1037.5--5647 & LS 1698  &B0V-IIIe  &860    &1.0$\pm$0.1	&4.7$\pm$0.4    
& 0.5  & 20     
& $\leq$ 65 \\
RX J0440.9+4431 & BSD 24--491&B0V-IIIe &202    &1.5$\pm$0.4	&$\sim$4.5    & 
0.03 & 6      
& $\leq$ 100 \\
\hline
EXO 2030+375& & Be   &41.7   &1.3-1.8$\pm$0.01$^a$ &5-19$\pm$0.5$^a$&100 & 1000  
 & 200 \\
4U0115+63 & V635 Cas    & O9e    &3.6    &0.36$\pm$0.01 &7.5$\pm$0.2    & 30   & 
200    & 
110 
\\
\hline
\end{tabular}
\end{center}

$a$ dependent on luminosity; higher values correspond to higher luminosity \\
$b$ energy range 1-20 keV except that of LS I+61 235 which is 0.5-10 keV \\
$c$ L$_{\rm x, max}$/L$_{\rm x, min}$
\end{table*}

Recently, Haberl et al. (1998) and Haberl, Angelini \& Motch (1998)
have suggested that LS I +61 235/RX J0146.9+6121 may be another
persistent BeXRB. This source is the pulsar with the longest spin
period ($\sim$ 1400 s). With two such systems one can look for more
similarities in an attempt to establish the existence of a distinct
subclass of BeXRBs. Comparing the X-ray observational characteristics
of X Per and LS I +61 235 we find that they both share the following
properties:

\begin{itemize}

\item[-] Long pulse periods, 837s for X Per and 1404s for LS I +61
235

\item[-] Persistent, low luminosity ($\leq$10$^{34-35}$ erg s$^{-1}$) X-ray 
emission

\item[-] Low cut-off energy derived from the spectral fitting,
compared to the typical value of 10-20 keV found in transient X-ray
pulsars -- 2.2 keV for X Per (Schlegel et al. 1993) and 4 keV for LS I
+61 235 (Haberl, Angelini \& Motch 1998)
	
\item[-] Absent or very weak iron line at 6.4 keV, indicative of only 
small amounts of material in the vicinity of the neutron star

\item[-] Low X-ray variability. Flat lightcurves with rare and unpredictable 
increases in flux by a factor of $\leq$ $\sim$10

\item[-] No dependence of the X-ray spectrum on intensity

\end{itemize}

If we use these general characteristics as defining a subclass of
BeXRBs, we find that the newly discovered systems LS 1698 and BSD
24--491 are also potential members.  Both have been detected every
time they have been observed, indicating persistent X-ray emission,
and their X-ray flux is relatively low compared to that of other
accreting pulsars when they are active.  Furthermore, both have long
spin periods ($\sim$860 s for LS 1698, $\sim$202.5 s for BSD 24--491)
and low cut-off energies of 6.2 (4.7) and 1.9 (4.5) keV respectively -
the values in brackets are obtained when an iron line is included in
the fit, but this is not statistically required, again as in the cases
of X Per and LS I +61 235.  We have also shown in Sect \ref{HR} that
the HRs did not change significantly throughout the observations,
indicating that the X-ray spectrum is insensitive to changes in the
count rate.  This absence of spectral changes with intensity was also
seen in LS I +61 235 by Haberl, Angellini \& Motch (1998).  In
contrast, transient BeXRBs show, in general, a positive correlation
between HR and intensity, as seen in LS 992 (Reig \& Roche 1999) or
EXO 2030+375 (Reig \& Coe 1998).  It is worth noting that displaying
just one of this characteristics is not enough to qualify as a member
of the group, but the combination is suggestive of a real association.
For example, the X-ray spectrum of LS 992 also presents a low cut-off
energy (4.7 keV), but it does not share any of the other X-ray
properties (Reig \& Roche, 1999). The X-ray properties of these
persistent BeXRBs and some examples of the transient BeXRBs are
compared in Table 3.

It therefore seems that we now have a growing subclass of BeXRBs,
characterised by persistent, low-luminosity X-ray emission and slowly
rotating pulsars. A possible model for these systems is that of a
neutron star orbiting a Be star in a relatively wide orbit, accreting
material from only the low density outer regions of the circumstellar
envelope. Sporadic ejection of the disc itself, however, will result
in the ejected material and the neutron star interacting, producing a
flare in the X-ray lightcurve. The increase in flux during these
flares is typically less than a factor $\sim$10 (Roche et al. 1993),
although further work needs to be done on characterising the range of
X-ray variabilities observed in the whole range of BeXRBs. Note that
the expelled material can reach the neutron star at any orbital phase
(not necessarily at periastron) and that flares need not be
periodic. Thus, the combination of little material being accreted plus
slow rotation plus possibly a relatively weak magnetic field makes
these systems weak but persistent pulsing sources.

Due to the low accretion rate outside of these rare events (e.g.  X
Per has undergone only one observed episode of X-ray flaring activity
in almost 30 years, associated with major changes in the circumstellar
environment, Roche et al. 1993 and 1997; LS I +61 235 only 2 flares in 13
years, Haberl, Angelini \& Motch 1998), the neutron star may not be
spinning at the equilibrium period, and will thus not follow the
relationship predicted from the Corbet diagram of P$_{spin}$ v
P$_{orbit}$ (Corbet 1986; Waters \& van Kerkwijk 1989).  However, a
long orbital period is expected, to account for the presumed distance
of the neutron star from the envelope.

Of these persistent systems, only X Per has been the subject of
detailed, long-term study. In this case, whilst many of the physical
characteristics of the system have been elucidated (e.g. stellar
parameters by Lyubimkov et al. (1997), disc parameters by Telting et
al. (1998), photometric and polarimetric behaviour by Roche et
al. (1997) etc.), there are still doubts about fundamental parameters
such as the orbital period. None of the proposed members of this
subclass has an orbital period determination, and so their position on
the Corbet diagram cannot be ascertained. Future work should therefore
focus on a search for orbital variations in these objects.

\section{Conclusions}

We have discovered two new Be/X-ray pulsars which appear to display
X-ray characteristics in common with the proposed subclass of
persistent BeXRBs currently consisting of X Per and LS I +61 235. If
the association of LS 1698 and BD 24--491 with these other objects is
correct, the sample of persistent BeXRBs has doubled, and further
observations of all these sources is encouraged to study the
properties of this subclass. Currently, only X Per has been studied
extensively for any duration, but the existence of similar systems
provides us with probes of the behaviour of neutron stars in low
density accretion regimes. A possible model for these systems is that
of a neutron star orbiting a Be star in a relatively wide orbit,
accreting material from only the low density outer regions of the
circumstellar envelope. Small increases in X-ray luminosity (e.g. of
order 10 times the quiescent flux) may result from mass ejection
episodes where the neutron star is temporarily in a region of
increased matter density.

\subsection*{Acknowledgments}

P. Reig acknowledges support via the European Union Training and
Mobility of Researchers Network Grant ERBFMRX/CP98/0195. We thank
I. Negueruela, D. Beckham, B. Scouza and V. Adams for useful discussions.

\bsp


\begin{thebibliography}{99}

\bibitem{} Corbet R.H.D., 1986, MNRAS, 220, 1047.
\bibitem{} Haberl F., Angelini L., Motch C., White N.E., 1998, A\&A, 330, 
189.
\bibitem{} Haberl F., Angelini L., Motch C., 1998, A\&A, 335, 587.
\bibitem{} Hutchings J.B., Cowley A.P., Crampton D., Redman
R.O., 1974, ApJ, 191, L101.
\bibitem{} Jahoda K., Swank J.H., Stark M.J., Strohmayer T., Zhang W., Morgan
E.H., 1996 {\it EUV, X-ray and Gamma-ray Instrumentation for Space
Astronomy VII}, O.H.W. Siegmund \& M.A. Gummin eds., SPIE 2808, 59, 1996.
\bibitem{} Lyubimkov L.S., Rostopchin S.I., Roche P., Tarasov
A.E., 1997, MNRAS, 286, 549.
\bibitem{} Markert T.H., Laird F.N., Clark G.W. et al., 1979, ApJS,
39, 573.
\bibitem{b2} Motch C., Haberl F., Dennerl K., Pakull M., Janot-Pacheco E., 
1997, A\&A, 323, 853. M97
\bibitem{} Negueruela I.N., 1998, A\&A in press.
\bibitem{} Reig P., Coe M.J. 1998, MNRAS, 294, 118.
\bibitem{} Reig P., Roche P., 1999, MNRAS, in press.
\bibitem{} Roche P., Coe M.J., Fabregat J., et al. 1993, A\&A, 270, 122.
\bibitem{} Roche P., Larionov, V.M., Tarasov, A.E., et al., 1997, A\&A,
322, 139. 
\bibitem{} Ryter C., Ceasarsky C.J., Audouze J., 1975, ApJ, 198, 103.
\bibitem{} Schlegel E.M., Serlemitsos P.J., Jahoda K., et al. 1993, ApJ 407, 
744.
\bibitem{} Telting J.H., Waters L.B.F.M., Roche P., et al., 1998,
MNRAS, 296, 785.
\bibitem{} van Paradijs J., 1995, in Lewin, van Paradijs \&
van den Heuvel, {\em X-ray Binaries}, Cambridge University Press.
\bibitem{} Waters L.B.F.M. \& van Kerkwijk M.H., 1989, A\&A, 223, 196


\end{thebibliography}
\end{document}